\documentclass[showpacs,preprintnumbers,amsmath,amssymb,nofootinbib]{revtex4}

\usepackage{graphicx}
\usepackage{color}
\usepackage{latexsym}
\newcommand{\blue}{\textcolor{blue}}
\newcommand{\red}{\textcolor{red}}
\newcommand{\green}{\textcolor{green}}

\newcommand{\be}{\begin{equation}}
\newcommand{\ba}{\begin{eqnarray}}
\newcommand{\ee}{\end{equation}}
\newcommand{\ea}{\end{eqnarray}}
\newcommand{\vphi}{\varphi}
\newcommand{\barr}{\begin{array}}
\newcommand{\ear}{\end{array}}

\begin{document}

\preprint{CERN-PH-TH/2004-106}  

\vspace{3cm}

\title{Moving Five-Branes and Membrane Instantons in Low Energy Heterotic M-Theory}

\author{Beatriz de Carlos}
\email{B.de-Carlos@sussex.ac.uk}
\altaffiliation[also at ]{Department of Physics CERN, Theory Division, 1211 Geneva 23, Switzerland}
\affiliation{Department of Physics and Astronomy, University of Sussex, Brighton, BN1 9QJ, UK}
\author{Jonathan Roberts}
\email{j.roberts@sussex.ac.uk}
\affiliation{Department of Physics and Astronomy, University of Sussex, Brighton, BN1 9QJ, UK}
\author{Yaiza Schm\"ohe}
\email{y.schmohe@sussex.ac.uk}
\affiliation{Department of Physics and Astronomy, University of Sussex, Brighton, BN1 9QJ, UK}
\date{\today}
\begin{abstract}

We study cosmological solutions in the context of 4-dimensional low energy 
Heterotic M-theory with moving bulk branes. 
In particular we present  non-trivial, analytic 
axion solutions generated by new symmetries of the full potential-free action which are 
similar to 'triple axion' solutions found in Pre-Big-Bang (PBB) cosmologies. 
We also consider the presence of a  non-perturbative superpotential, for which we 
find cosmological solutions with and without  a background perfect fluid. 
In the absence of a fluid the dilaton and the $T$-modulus go to the
potential-free solutions at late time,  while the moving brane tries to avoid colliding with the 
boundary and stabilize within the bulk. 
When the fluid is included, we find that the real parts of the fields track its behaviour
and that the moving brane gets stabilized at the middle point between the boundaries. In this latter case we 
can make analytic approximations
for the evolution of the fields whether or not axions are included and we consider 
the possibility of this set up being a realization of the
quintessential scenario.

\end{abstract}

\maketitle

\section{INTRODUCTION}

In recent years, there has been growing interest in the role played by
extra dimensions in Early Universe cosmology. In particular, a lot of work has been devoted
to studying the so called brane-world cosmology, where the Universe is a 4-dimensional extended
object within a higher-dimensional space~\cite{ Lukas:1998yy,Lukas:1998qs, Randall:1999vf, Brandle:2000qp}.  
One particular, interesting, possibility is to consider moving 
branes~\cite{Dvali:1998pa,Khoury:2001wf, Steinhardt:2001vw, Alexander:2001ks, Burgess:2001fx, Kallosh:2001du, Kehagias:1999vr, Copeland:2001zp, Copeland:2002fv} . In this paper we will look at 
moving bulk brane cosmologies within a well defined M-theory context.

To be more explicit, our starting point is the 5-dimensional effective action that has been  derived 
by compactifying 11-dimensional heterotic M-theory~\cite{Horava:1996qa,Horava:1996ma} on a
Calabi-Yau 3-fold~\cite {Witten:1996mz,Horava:1996vs,Lukas:1998fg,Lukas:1998hk,Lukas:1998yy}. The
resulting theory is an explicit realisation of a brane
world scenario and consists of a 5-dimensional N=1 supergravity theory
on the orbifold $S^1/Z_2$ coupled to two N=1 4-dimensional theories at the
orbifold fixed points~\cite{Lukas:1998hk,Lukas:1998yy, Ellis:1998dh,Lukas:1998tt,Brandle:2001ts}.
It has been shown~\cite{Witten:1996mz,Lukas:1998hk,Lukas:1998yy,Lukas:1998qs,Derendinger:2000gy,Brandle:2001ts} 
that one is free to add M-theory 5 branes into
the 11-dimensional theory. Upon compactification, these branes appear
as 3-branes in the 5-dimensional bulk and are free to move along the
orbifold direction. To simplify the problem, we will use the
4-dimensional effective action obtained~\cite{Brandle:2001ts,Derendinger:2000gy} by reduction on the 5-dimensional 
BPS domain wall vacuum solution.

The minimal version of this set up contains three chiral
superfields, namely the dilaton $S$,
the universal $T$ modulus and the modulus $Z$, which specify the
position of the moving brane in the orbifold direction. These
superfields are constructed from six real scalar fields, three of which have a geometrical interpretation in 
terms of the higher dimensional theory. 

This scenario has been studied, in the absence of a potential, in Ref.~\cite{Copeland:2001zp}, where, it was shown
that the axionic components of the chiral superfields could be set to zero.
Then two of the  remaining component fields, $\vphi$ (measuring the Calabi-Yau volume) and the real part of the 
$T$ modulus, $\beta$ (measuring the orbifold size), behave asymptotically as rolling radii
(RR) fields~\cite{Mueller:1990rr}, while the bulk brane, $z$, moves a finite distance to generate a transition 
between the early and late time rolling radii solutions. These solutions are familiar from Pre-Big-Bang (PBB) 
cosmology and provide a framework for negative time branch cosmology within Heterotic M-Theory.  

In this paper, we will go further and take into account both new symmetries of the potential-free
action, in order to generate new axion-dependent solutions, and also
non-perturbative 
effects~\cite{Becker:1995pot,Lukas:1999kt,Harvey:1999as,Lima:2001jc, Lima:2001nh,Moore:2000fs,Curio:2001qi}. 
In particular, 
we will introduce a superpotential due to M2 instantons stretching between the boundaries
and the moving brane, and study the effects of the scalar
potential coming from this superpotential in the evolution of the
moduli.

The outline of the paper is as follows. In the next section, we will
explain our scenario in more detail, motivate the introduction of
this particular superpotential and present the  action we will work with and its symmetries.

In section~\ref{sec:new_sols}, we will show how, in the absence of a potential,
these symmetries generate new solutions, with non-trivial axion dynamics, from the ones found in 
Ref.~\cite{Copeland:2001zp}. These solutions are valid on either time branch but might be particularly important 
in generating scale invariant isocurvature perturbations in a similar fashion to the PBB scenario.

In section~\ref{sec:jon_sols}, we see that we can still consistently truncate off the axions and investigate the 
behaviour of $\vphi$, $\beta$ and $z$ in the presence of the membrane potential on the negative time branch. 
We find that,  at late times, the first two fields exhibit a subset of the late time rolling radii behaviour of the 
potential free case, but with the potential acting to prevent a bulk-boundary brane collision and possibly 
providing a way of transferring isocurvature perturbations into adiabatic ones (analogously to what happens, for 
example, in Ref.~\cite{DiMarco:2002eb}).

We will, in section~\ref{sec:yaiza}, introduce a background fluid into the system, as required for a  realistic 
model of the Universe evolving in the positive time branch. Firstly we will study its effect on the real parts 
of the  moduli by setting the axions to their vacuum values. 
We will see how the fields tend to scale with the background fluid,
allowing us  to find a set of analytic solutions in different limiting
cases.  We also consider the possibility of one of these moduli being the 
quintessence field that might be accelerating the Universe today.
Finally, we will include the axions' evolution towards their minima in the analysis and we will show
how this will not affect the late scaling behaviour of the real parts of the moduli.

We present our conclusions and future work in section~\ref{sec:conclusions}.  


\section{SET UP} 
\label{sec:setup}

Our 5-dimensional brane-world scenario comes from the compactification on a
Calabi-Yau three fold \cite{Lukas:1998yy,Lukas:1998tt,Witten:1996mz,Horava:1996vs,Lukas:1998fg,Lukas:1998hk} 
of the low-energy limit of 11-dimensional Horava-Witten theory~\cite{Horava:1996qa, Horava:1996ma}
 that is, 11-dimensional supergravity on the orbifold $S^1/Z_2\times M_{10}$ coupled to two
 10-dimensional $E_8$ Super-Yang-Mills theories, one residing on each of the two
 10-dimensional orbifold fixed planes. 
 This compactification leads \cite{Lukas:1998hk,Lukas:1998yy, Ellis:1998dh,Lukas:1998tt,Brandle:2001ts} in the bulk, 
to a 5 dimensional N=1 gauged supergravity on the orbifold $S^1/Z_2\times M_{4}$ coupled to two N=1 gauged theories on
 the now 4-dimensional orbifold fixed planes on which two 3-branes carry the
 so called observable and hidden sectors.

 In addition to this minimal set-up, we are free to include M-theory
 5-branes in the vacuum of the 11-dimensional theory \cite{Lukas:1998hk, Derendinger:2000gy}. Upon compactification two
 of their dimensions wrap holomorphic two cycles within the Calabi-Yau three-fold,
 leaving effective 3-branes in the 5 dimensional bulk. 
These additional 3-branes are
 transverse to the orbifold direction and have the interesting
 property of being able to move along that direction. Each of
 them will carry an additional N=1 supersymmetric theory. For the purpose of
 our paper, we will restrict ourselves to the inclusion of only one additional 3-brane. 
 The charges on the visible and hidden orbifold planes and the three
 brane, respectively $q_i$ (i=1,2,3), are quantised in units of
 $\epsilon_0$ and will satisfy the cohomology condition~\cite{Lukas:1999nh}
\begin{equation}
\sum_{i=1}^{3}q_{i}=0 \;\;,
\label{eq:cohomology}
\end{equation}
which follows from anomaly cancellation in the 11-dimensional theory. For the
compactification to preserve 4-dimensional supersymmetry, $q\equiv
q_3>0$ \cite{Brandle:2001ts}.
 
Practically, we will be working in the context of a 4-dimensional effective
 theory which arises from further reduction of the 5-dimensional theory on the
 domain-wall vacuum solution \cite{Derendinger:2000gy,Lukas:1998yy}. The simplest version of the D=4, N=1 action contains
 three chiral superfields,
 namely the dilaton $S$, the universal $T$ modulus and $Z$ which
 specifies the position of the additional brane in the orbifold
 direction. In terms of the underlying component fields, these
 superfields can be written as \cite{Brandle:2001ts}
\begin{eqnarray}
& &S = e^{\vphi}+qz^2e^{\beta}-2i(\sigma-qz^2\chi) \;\;, \nonumber\\
& &T = e^{\beta}+2i\chi  \;\;, \label{eq:compfields}\\
& &Z = e^{\beta}z-2i(\zeta-z\chi) \;\;, \nonumber
\end{eqnarray}
where $\vphi,\beta, z, \sigma, \chi, \zeta$, are all real scalar
fields, with $\vphi,\beta, z$ having 
a geometrical interpretation in terms of the higher-dimensional
theory. The field $\vphi$ comes from the 5-dimensional dilaton, a part of the universal D=5
hypermultiplet, and measures the size of the internal Calabi-Yau three-fold
averaged over the orbifold, so
that the Calabi-Yau volume, with respect to a fixed reference volume $v$, is given by $ve^{\vphi}$. The dilatonic 
axion, $\sigma$, will be the other zero mode surviving from the universal
hypermultiplet. The $\beta$ field corresponds to the zero mode of the
(55)-component in the $D=5$ metric and measures the size of the
orbifold (more precisely the orbifold size will be $\pi \rho
e^{\beta}$, where $\pi\rho$ is a fixed reference size). Its associated axion, $\chi$, is related to the vector field
in the 5-dimensional gravity multiplet. The field $z$ originates from the world
volume of the additional 3-brane and specifies the position of the
brane in the orbifold direction; it is normalised as $z\in[0,1]$, so
that $z=0$ ($z=1$) will correspond to the observable (hidden) orbifold fixed
plane. The associated axion $\zeta$, originates from the additional
3-brane and is related to the self-dual two-form of the underlying
5-brane.

The K\"ahler potential which governs the dynamics of these three chiral superfields, has been computed in 
Refs~\cite{Derendinger:2000gy,Brandle:2001ts} and is given by
\begin{equation}
K=-\ln\left(S+\overline{S}-q\frac{(Z+\overline{Z})^2}{T+\overline{T}}\right)-3\ln(T+\overline{T}) \;\;,
\label{eq:kahler}
\end{equation}
at lowest order in $\epsilon$, the strong coupling expansion parameter
\begin{equation}
\epsilon=qe^{\beta-\vphi} \;\;,
\label{eq:epsilon}
\end{equation}
which measures the size of the string-loop corrections to the 4-dimensional effective
action or the strength of the Kaluza-Klein excitations in the orbifold
direction. From the 5-dimensional point of view we could understand it as the warping of the domain wall
vacuum solution in the orbifold direction, or geometrically, the relative size of the orbifold and the Calabi-Yau. 
 
Considering the range of validity of our effective theory allows us to place constraints on regions of superfield 
(or, equivalently, component field) parameter space in which our solutions can be trusted. As stated above, 
the K\"ahler potential, eq.(\ref{eq:kahler}), is computed to first order in the stong coupling parameter, $\epsilon$. 
Thus our model will only  be valid in the weakly coupling regime, $\epsilon\ll 1$, where
\be
S_R > \frac{qZ_R^2}{T_R}+qT_R ~~ \big(\longleftrightarrow e^\vphi > qe^\beta\big) \;.\label{eq:parameterspace1}
\ee
In addition ensuring that $\alpha'$ corrections remain small, requires
\be
T_R \geq \frac{1}{8\pi} ~~ \big(\longleftrightarrow \beta \geq -\ln(8\pi)\big)\;.
\label{eq:parameterspace2}
\ee
Finally as we have not modelled bulk brane - boundary brane collisions our effective theory will break down if the 
bulk 3-brane strikes a boundary, i.e.
\be
Z_R \in[0,T_R]  ~~ \big(\longleftrightarrow z\in[0,1] \big)\;.
\label{eq:parameterspace3}
\ee
Subject to these considerations we now derive our effective theory.

The four-dimensional N=1 supergravity action, is of the form
\begin{equation}
S=\frac{-1}{2\kappa_P^2}\int \sqrt{-g} \left(\frac{1}{2}R+K_{i\bar j}\partial_{\mu}\Phi ^{i}\partial^{\mu}{\bar{\Phi}^{\bar{j}}}+V\right)d^4x \;\;,
\label{eq:action}
\end{equation}
where $ K_{i\bar j}=\frac{\partial^{2}K}{\partial\Phi^{i}\partial\bar\Phi^{\bar j}}$
is the K\"ahler metric (given in Appendix~\ref{sec:Kahler}); $\Phi^{i}=(S,T,Z)$ are the three complex chiral 
superfields; V($\Phi$) is the scalar
potential and $\kappa_P$ is the 4-dimensional Newton constant which, in
terms of higher dimensional quantities, can be expressed as
\begin{equation}
\kappa_{P}^2=\frac{\kappa^2}{2\pi\rho v}=8\pi G_N \;\;,
\label{eq:kappa}
\end{equation}
where $\kappa$ is the 11-dimensional Newton constant.

Initial studies have been carried out upon the dynamics of this system
in the absence of a potential. The full 4-dimensional solutions were found in Ref.~[5]
and some aspects of the 5-dimensional solutions were discussed in Ref.~[4]. In this
article we will go beyond the perturbative approach, where $V=0$, and
consider non-perturbative effects, which induce an effective scalar potential whose most general form for 
4-dimensional $N=1$ supergravity is 
\begin{equation}
V =
e^{K}(K^{i\overline{j}}D_iW\overline{D_{j}W}-3W\overline{W}) \;\;.
\label{eq:potential}
\end{equation}
Here $K^{i\overline{j}}$ is the inverse K\"ahler metric and $D_iW=\partial_{i}W+\frac{\partial K}{\partial \Phi^{i}}W$ 
is the K\"ahler covariant derivative acting on the superpotential. 

There are a number of non-perturbative effects we could consider 
\cite{Becker:1995pot,Lukas:1999kt,Harvey:1999as,Lima:2001jc,Lima:2001nh,Moore:2000fs}, each contributing to the 
superpotential for the chiral superfields. Possible effects include \cite{Moore:2000fs}: gaugino condensation, 
M2-brane instantons stretching between the two boundaries, an M5-brane wrapping the whole Calabi-Yau and M2 
instantons that stretch between the nine-boundaries and the five-branes in the higher
dimensional theory. This last contribution is amongst the dominant, therefore we
will study its effect in our theory while neglecting the others. Specifically, in our model with
 just one additional 3-brane, the superpotential takes the form \cite{Moore:2000fs}
\begin{equation}
W_{M2,M5} = h(e^{-2\pi qZ}+e^{2\pi q(Z-T)}) \;\;,
\label{eq:superp}
\end{equation}
where $h$ is of the order of the Planck mass and, for practical purposes, we will use $h=1$. The first term of 
eq.~(\ref{eq:superp}) will correspond to the membrane stretching between the visible boundary and the bulk 3-brane, 
while the second one represents the M2 stretching from the bulk 3-brane to the hidden
orbifold fixed plane.

 We can now compute the scalar potential, eq.~(\ref{eq:potential}), for this particular case, which can be written as 
\begin{equation}
V=V_{0}(\phi_R)[A(\phi_R)+B(\phi_R)\cos\theta(\phi_I)] \;\;,
\label{eq:scalarpot}
\end{equation}
where $\phi_R$=$(S_R,T_R,Z_R)$ are the real parts of the chiral superfields, whereas
$\phi_I$=$(S_I,T_I,Z_I)$ represent the imaginary parts. It is
  worth noting that only two of the imaginary parts of the superfields appear
  in the cosine term,
\be
\theta(T_I,Z_I)=2\pi q(2Z_I-T_I)\longleftrightarrow\theta(\chi,\zeta,z)=2\pi q(-4\zeta+4z\chi-2\chi) \;\;,
\label{eq:theta}
\ee
whereas the other terms in the potential depend solely on the real part
 of the superfields in a polynomial and exponential way. The
 full potential in terms of both superfields and component fields is given 
in Appendix~\ref{sec:app_pot}.

The action described in eq.~(\ref{eq:action}) has the following symmetries

{\em Boundary-exchange symmetry}

The full action is invariant under following variations in the superfields, 
\ba
S&\longrightarrow& S+q(T-2Z) \;\;, \nonumber\\
T&\longrightarrow& T \;\;, \label{eq:csfsym} \\
Z&\longrightarrow& -Z+T \;\;.\nonumber
\ea
We can see from the previous equation that  $Z=T/2$ is a fixed point under this symmetry, and this fact will be 
exploited when making our analytic approximations in section~\ref{sec:yaiza}. The geometrical interpretation of 
this symmetry becomes evident if we recast it in terms of the component fields. Inserting eq.~(\ref{eq:compfields}) 
into eq.~(\ref{eq:csfsym}) we find 
\ba
\vphi\rightarrow\vphi & , & \sigma \rightarrow \sigma - 2 q \zeta \;\;, \nonumber \\
\beta\rightarrow\beta & , & \chi \rightarrow \chi \;\;, \\
z\rightarrow(1-z) & , & \zeta \rightarrow -\zeta \;\;, \nonumber 
\ea 
and we see that our symmetry exchanges boundary branes.

The K\"ahler potential, eq.~(\ref{eq:kahler}), will always be
invariant under this transformation however the superpotential, eq.~(\ref{eq:superp}), is only invariant as we 
have chosen $h$, the prefactor in front of the exponentials, to be the same for both membrane instantons. 

{\em Axionic symmetries}

When we write the action, eq.~(\ref{eq:action}), in terms of the component fields we find, in the absence of a 
potential, two trivial shift symmetries in the axions
\be
\sigma\rightarrow\sigma, ~\chi\rightarrow\chi, ~\zeta\rightarrow\zeta+c_1 \;\;,
\label{eq:sym1}
\ee
and
\be
\sigma\rightarrow\sigma+c_2, ~\chi\rightarrow\chi, ~\zeta\rightarrow\zeta \;\;,
\label{eq:sym2}
\ee
and a third, less obvious, symmetry which mixes the axions with $z$, the bulk brane position
\ba
\sigma &\rightarrow& \sigma + qc_3z^2, \nonumber \\ 
\chi &\rightarrow& \chi+c_3,\label{eq:sym3}\\
\zeta &\rightarrow& \zeta+c_3z\nonumber\;\;.
\ea
This can be understood as the reduction to 4 dimensions~\footnote{The associated 5 dimensional symmetries are 
given in Ref.~\cite{Brandle:2001ts}.} of the well known (e.g. Ref.~\cite{Brandle:2001ts}) gauge symmetries of 
the effective $D=11$ action for Horava-Witten theory coupled to a five brane.

We saw in eqs.~(\ref{eq:scalarpot}) and (\ref{eq:theta}), that only two axions appear in the potential and that 
they are contained in its $\cos\theta$ term. Therefore, although $\theta$ is not invariant under the first, 
$\theta\rightarrow\theta +c_{1}$, or third, $\theta\rightarrow\theta +c_{3}$, symmetries, the potential's dependence 
on $\cos\theta$ means that we only break the full symmetries to discrete ones. In other words, $c_1=\pm n/q$ in 
eq.~(\ref{eq:sym1}) whereas $c_3=\pm n/2q$ in eq.~(\ref{eq:sym3}). The second symmetry, eq.~(\ref{eq:sym2}), is 
unaffected as $V$ is independent of $\sigma$.

\section{NEW POTENTIAL FREE COSMOLOGICAL SOLUTIONS} 
\label{sec:new_sols}

In Ref.~\cite{Copeland:2001zp} it was shown that, when considering
solely the kinetic terms, one can consistently truncate the component field
action by setting $\sigma$ and $\zeta$ to constants and $\chi$ to zero
to leave the maximally truncated moving brane action. We can now use
the third axion symmetry, eq.~(\ref{eq:sym3}), to generate a new, more general, class of analytic solutions, 
which include non trivial dynamics for the axions $\zeta$ and $\sigma$, from the solutions found in 
Ref.~\cite{Copeland:2001zp}. 

Before we generate our new solutions let us briefly discuss the solutions found in Ref.~\cite{Copeland:2001zp}. 
These will be relevant both in the context of our new axionic solutions and in certain limits of the potential 
driven system. They represent the solutions to the M-theory 4-dimensional effective action, explicitly given by
\be
S= - \frac{1}{2\kappa_{P}^{2}} \int\sqrt{(-g)}\left(\frac{R}{2} + \frac{1}{4} \partial_{\mu}\vphi\partial^{\mu}\vphi + 
    \frac{3}{4} \partial_{\mu}\beta\partial^{\mu}\beta + 
    \frac{q}{2} e^{\beta -\vphi}\partial_{\mu}z\partial^{\mu}z \right)d^{4}x \;\;.
\label{eq:min_action}
\ee 
The cosmological solutions in flat FRW space were found to be
\ba
\alpha &=& \frac{1}{3}\ln\left|t\right|+\alpha_0 \;\;, \nonumber \\
\beta &=& p_{\beta ,i}\ln\left|t\right|+  (p_{\beta ,f}-p_{\beta ,i})
   \ln\left(\left|t\right|^{-\delta}+1 \right)^{-\frac{1}{\delta}}+\beta_0 
     \;\;, \label{eq:min_sol} \\
\vphi &=& p_{\vphi ,i}\ln\left|t\right| +  (p_{\vphi ,f}-p_{\vphi ,i})
   \ln\left(\left|t\right|^{-\delta}+1\right)^{-\frac{1}{\delta}}+\vphi_0 
     \;\;, \nonumber  \\
z &=& d \left( 1 + \left|t\right|^{\delta} \right)^{-1} + z_0 \nonumber \;\;.
\ea
The initial and final expansion powers of the fields $\beta$ and $\vphi$ are subject to the constraint
\begin{equation}
 3p_{\beta ,n}^2+p_{\vphi ,n}^2=\frac{4}{3},~~ n=i,f  \;\;,
\label{eq:cons1}
\end{equation}
which defines an ellipse in the $(p_{\beta}, p_{\vphi,})$ plane, and are related by the linear map
\begin{equation}
 \left(\barr{c}p_{\beta ,f}\\p_{\vphi ,f}\ear\right) = P
 \left(\barr{c}p_{\beta ,i}\\p_{\vphi ,i}\ear\right)\; ,\qquad
 P = \frac{1}{2}\left(\barr{rr}1&1\\3&-1\ear\right)\; .
\label{eq:map1}
\end{equation}
This ellipse and linear map can be seen in figure~\ref{fig:RR} where we have used them to understand the evolution 
of the fields in our potential driven solutions. 

The power $\delta$ is given by $\delta = p_{\beta ,i}-p_{\vphi ,i}$ and can be fixed such that
\ba
\delta>&0& (-)~  {\rm timebranch} \;, \nonumber  \\
\delta<&0& (+)~  {\rm timebranch}  \nonumber \;.
\ea
Note that the initial position of the bulk brane, $z_0$, is unconstrained, however the distance it moves, $d$, is 
constrained by the condition 
\begin{equation}
\vphi_0-\beta_0 = \ln\left(\frac{2qd^2}{3}\right)\; .
\end{equation}
In the asymptotic limits $(t\rightarrow \pm 0, t\rightarrow \pm\infty)$, $z$ is a constant and the fields $\vphi$ and 
$\beta$ behave as freely rolling radii, with their early (late) time expansion parameters given by $p_{\vphi ,i}$ and 
$p_{\beta ,i}$ ($p_{\vphi ,f}$ and $p_{\beta ,f}$)~\footnote{The parameters in both limits are constrained by 
eq.~(\ref{eq:cons1}) however, at early times, they can only be on the part of the ellipse (\ref{eq:cons1}) where  
$p_{\beta ,i}-p_{\vphi ,i} > 0$, while at late times the fields satisfy  $p_{\beta ,f}-p_{\vphi ,f} < 0$. These 
negative time branch constraints get reversed for the positive time branch.}. In the intervening period the 3-brane 
moves significantly, leading to a more complicated evolution of the fields, but ultimately generates a transition from 
the early to late time rolling radii regimes. It is worth noting that the bulk brane does not necessarily collide with 
a boundary, it starts at $z_0$ and moves a fixed distance $d$, to become constant again at $z_0 + d$. Thus, 
provided $z_0+d \in [0,1]$, there will be no collision with a boundary.

Now when we generate the axionic solutions from our new symmetry eq.~(\ref{eq:sym3}), we see that their dependence 
on the moving brane ensures that the axions remain constant in the asymptotic limits, with the bulk brane movement 
generating a transition between these positions. Applying the non trivial symmetry, eq.~(\ref{eq:sym3}), to the 
solution for $z$, eq.~(\ref{eq:min_sol}), we see that the behaviour of the axions is explicitly given by
\ba
\sigma &=&  qc_3d^2\left( 1 + \left|t\right|^{\delta} \right)^{-2}+ qc_3z_0d\left( 1 + \left|t\right|^{ \delta} \right)^{-1} + \sigma_0  \;, \nonumber\\
\chi&=&c_3 \;, \label{eq:axion_sols} \\
\zeta &=&  c_3 d\left( 1 + \left|t\right|^{\delta} \right)^{-1} + \zeta_0  \;. \nonumber 
\ea
It is worth noting that, as in the case of the minimal solutions, these new ones have similarities to solutions found 
in the PBB literature. In Ref \cite{Copeland:2001zp} it was noted that the minimal moving brane solutions presented 
above, eqs. (\ref{eq:min_sol}), are similar to the dilaton-moduli-axion solutions \cite{Copeland:1994vi} of PBB 
cosmology \cite{Gasperini:1993em}, with the moving brane playing the role of the axion. Our new solutions, 
eqs.~(\ref{eq:axion_sols}), in combination with the unaffected minimal solutions, eqs. ~(\ref{eq:min_sol}), are 
themselves similar to the dilaton-moduli-'3 axion' solutions derived from an SL(3,R) invariant action in 
Ref.~\cite{Lidsey:1999mc}. This suggests that, as in the PBB case, we would expect that our axions will provide us 
with a scale invariant isocurvature perturbation.

In summary, although we generate non-trivial axion dynamics, the behaviour of the scale factor and the other, 
non-axionic, fields remains unchanged in this more general class of solutions. What we see is that the moving brane 
not only causes a transition in the moduli fields, as described above, but also causes the axions to evolve smoothly 
from one constant value to another. It should be noted that, as in the axion free case, the strong coupling 
parameter $\epsilon$ grows asymptotically and the effective theory breaks down. At this point we can no longer 
say whether $z$ or the axions remain constant.

\section{Including non-perturbative effects}
\label{sec:jon_sols}

We now return to the main topic of this paper, i.e. the inclusion of a non-perturbative superpotential into the 
dynamics of the 4-dimensional effective theory of heterotic M-theory. As discussed above we will only consider the 
effect of two membrane instantons, one stretching from each boundary to a single (moving) bulk brane. In this 
section we will concentrate on the component fields and their potential, as
well as on simplifying  the action as much as possible.

From eq.~(\ref{eq:scalarpot}) it is clear that all dependence on the axions is contained in
${\rm cos} (\theta(\phi_I))$  and, therefore, we will have  a minimum of  the potential along the imaginary 
directions for either $\theta=0$ or $\theta=\pi$. So, in principle, one could integrate out 
the imaginary fields and work with a simplified action. However, given that 
$V_{0}(\phi_R)B(\phi_R)$ is a time dependent function of the real fields $\vphi$, $\beta$ and $z$ 
(or, equivalently, $S_R$, $T_R$ and $Z_R$), we need to ensure that, if the axions are initially set to a 
minimum of the potential, this point remains a minimum. That is, we need to ensure that $V_{0}B$, has the same 
sign throughout the evolution of our system. This is equivalent, see Appendix~\ref{sec:app_pot}, to ensuring that 
the following function does not change sign
\be
M=(3-24\pi^{2}qe^{\beta+\vphi}+16\pi^{2}q^{2}e^{2\beta}z(z-1)+12\pi qe^{\beta}) \label{eq:m}\;\;.
\ee
Independently of truncating the axions the validity of our effective
theory already restricts the available $(\phi, \beta)$ space, as shown
in eqs.~(\ref{eq:parameterspace1})-(\ref{eq:parameterspace3}).
Then to truncate the axions our two possible choices are\\ 
{\bf 1 - $M<0$}. In the weak coupling regime, this condition is satisfied provided
\be
\vphi > \ln\left(\frac{e^{-\beta}}{8\pi^2q} + \frac{1}{2\pi} \right) \;.
\ee
This leaves almost all of the remaining field space in which we can consistently truncate the axions.\\
{\bf 2 - $M>0$}. Reversing our choice leaves a very small region of
$\vphi$, $\beta$ space in which we can consistently truncate the
axions within a reliable 4-dimensional effective theory. 

Therefore, in the remainder of this section we will concentrate on the first, larger area of parameter space in which 
we can safely set $\chi=0$, $\zeta=0$ (or $\pm n/4q$) and $\sigma={\rm const}$. Explicitly our action now reads
\be
S= - \frac{1}{2\kappa_{P}^{2}} \int\sqrt{(-g)}\left(\frac{R}{2} + \frac{1}{4} \partial_{\mu}\vphi\partial^{\mu}\vphi +  \frac{3}{4} \partial_{\mu}\beta\partial^{\mu}\beta + 
    \frac{q}{2} e^{\beta -\vphi}\partial_{\mu}z\partial^{\mu}z + V(\vphi,\beta,z) \right)d^{4}x \;\;,
\label{eq:the_action}
\ee
where $V=V_0(\phi_i)[A(\phi_i)+B(\phi_i)]$, with $\phi_i=(\vphi,\beta,z)$. 

\subsection{Equations of Motion}

We now look at the cosmological solutions of the system governed by our truncated action eq.~(\ref{eq:the_action}). 
We consider the simple case of homogeneous, time-dependent, fields in a spatial flat FRW space time. Explicitly our 
ansatz reads
\ba
 ds^{2} &=& -dt^{2}+e^{2\alpha(t)}d{\bf x}^{2} \; , \nonumber \\
\phi_{i} &=& \phi_{i}(t) \;\; . \nonumber  \label{eq:ansatz} 
\ea
As usual our solutions will have two branches, a negative  branch with $t,\dot{\alpha} <0$, and a future curvature 
singularity at $t=0_-$, and a positive branch  with $t,\dot{\alpha} >0$, and a past curvature singularity at $t=0_+$. 
In this section we will concentrate on the negative  branch while in section \ref{sec:yaiza}, when we include a 
perfect fluid, we will look at the positive one. 

In general the equations derived from our action cannot be solved analytically, and we will resort to numerical 
analysis. However we can sometimes make useful approximations. Inspecting the form of the potential given by 
eq.~(\ref{eq:real_pot}), we expect the dominant contributions to come from the double exponential in the $\beta$ 
direction,  given that the $\vphi$ and $z$ directions are only proportional to single exponentials. Thus the 
potential acts to expand the orbifold. This suggest that, unless the kinetic energy or friction terms prevent the 
fields from running down the potential, the system will evolve towards the $V=0$ solutions (see 
section~\ref{sec:new_sols}) at late times. 

In this context it is useful to parameterise the fields as RR fields, i.e. let
\ba
\vphi=p_{\vphi} \ln |t| + \vphi_{0} \;\;, \nonumber \\
\beta=p_{\beta} \ln |t| + \beta_{0} \;\;,
\label{eq:RR_approx}
\ea
with the RR parameters given by the expansion coefficients $p_{\beta}$ and $p_{\vphi}$. Using these
equations we can retrieve effective RR parameters from the numerical
solutions to describe the evolution of our fields. Recall that, in the
potential free case, the fields make a smooth transition from one RR
solution to another (defined by the map in eq.~(\ref{eq:map1})), which means that, although the expansion parameters 
are no longer constant when the bulk brane is moving, they evolve in a well-defined way from one value to another. 
Thus, provided the potential does not totally dominate the dynamics, the RR parameterisation of our numerical results 
will be a useful way to get a handle on the evolution of the moduli fields. 

It will also be useful to consider the shape of our potential in the regimes where the RR parameterisation is valid. 
When we substitute eq.~(\ref{eq:RR_approx}) into the potential eq.~(\ref{eq:real_pot}), neglect the constants and 
note that $z\in[0,1]$ in the regions where our theory is valid, we find that all the exponential terms have the 
same shape. Thus the form of our potential can be approximated by
\be
V(\vphi ,\beta ,z) \approx [(a|t|^{-3p_{\beta} -p_{\vphi}} + b|t|^{-p_{\beta}-p_{\vphi}} + c|t|^{-2p_{\beta}})e^{-d|t|^{p_{\beta}}}] \;\;,
\label{eq:RR_approx_pot}
\ee
where $a$, $b$, $c$, $d$ $\in R^{+}$.

For our numerical analysis it proves convenient to use $\alpha$, rather than $t$, as our time parameter, although we 
will still use the time derivative of the fields to define our velocities. Practically this means we will 
numerically integrate the following equations 
\ba
\dot{\vphi}^\prime &=& - 3\dot{\vphi} - \frac{1}{\dot{\alpha}}qe^{\beta -\vphi}\dot{z}^{2} - \frac{2}{\dot{\alpha}}V,_{\vphi} \;, \nonumber \\ 
\dot{\beta}^\prime &=& -3\dot{\beta} + \frac{1}{3\dot{\alpha}}qe^{\beta -\vphi}\dot{z}^{2} - \frac{2}{3\dot{\alpha}}V,_{\beta}  \;, \\ 
\dot{z}^\prime  &=&    - \frac{1}{\dot{\alpha}}(3\dot{\alpha}+\dot{\beta}-\dot{\vphi})\dot{z} - \frac{1}{q\dot{\alpha}}e^{\vphi -\beta}V,_{z} \;, \nonumber
\label{eq:jon_eom}
\ea
where $\dot{} = \frac{\partial}{\partial t}$, $\prime = \frac{\partial}{\partial\alpha}$ and all variables are 
functions of $\alpha$. In addition we find that the Friedmann equation gives
\be
\dot{\alpha}^2 = \frac{1}{3} (\frac{1}{4}\dot{\vphi}^{2} + \frac{3}{4}\dot{\beta}^{2} +  \frac{q}{2}e^{\beta -\vphi}\dot{z}^{2} + V)  \;,
\label{eq:Friedmann}
\ee
with $\dot{\alpha} = -\sqrt{\dot{\alpha}^2}$ as we are considering the negative time branch.

Within this framework we will not solve for the scale factor. This means that in order to retrieve the RR parameters 
from the numerics, we need to parameterise $\alpha = p_{\alpha} \ln |t|  + \alpha_{0}$. We can then use the Friedmann 
equation to find $p_\alpha$ ($p_\alpha = \dot{\alpha}/\dot{\alpha}^\prime$). Note that $p_{\alpha}\rightarrow (1/3)$ 
when the potential becomes negligible and the scale factor has the usual (1/3) power law evolution of a system governed 
predominantly by kinetic energy, i.e eq.~(\ref{eq:min_sol}).

\subsection{An explicit example}

To describe the effects of the potential on the dynamics of the system we present and discuss an explicit example. 
In order to highlight most of the key effects of including a potential we have chosen an example where both potential 
and kinetic terms are important. We will then comment on some features not found in this example to develop a more 
complete picture of the model.

In our example the potential plays an important role at early times without totally dominating the dynamics. Then, as 
it becomes less important, the system behaves similarly to the potential free case but punctuated by periods of more 
complex evolution, as the bulk brane approaches the boundary and is repelled by the potential.

We start with a contracting Calabi-Yau (C-Y), an approximately static orbifold and a slowly moving bulk brane at 
$\alpha=3$. Specifically we have
\ba
\vphi_{in}&=&3.1, ~~ \beta_{in}=1.0, ~~ z_{in}=0.3 \nonumber  \;\;, \\
\dot{\vphi}_{in}&=&0.001, ~~ \dot{\beta}_{in}=0.4\times10^{-7}, ~~ \dot{z}_{in}=0.6\times10^{-7} \;\;.
\ea
In the absence of a potential these initial conditions would lead to a very similar evolution to the explicit example 
given in Ref.~\cite{Copeland:2001zp}, i.e. the C-Y continues to contract while the orbifold and bulk brane remain 
almost static until $\alpha\approx0$. At this point the bulk brane moves appreciably to its new stable value causing 
the other fields ($\vphi$ and $\beta$) to make the transition to their late time RR values.

In this case the non negligible initial potential means that the fields do not behave as RR fields at early times. 
Instead the fields roll rapidly down the potential and go straight into the transition period where the bulk brane 
is moving appreciably and $\vphi$ and $\beta$ are evolving towards their late time solutions.  

After this initial, brief period, the potential is only important when the bulk brane approaches the boundaries, and 
the fields evolve as in the potential free case.
\begin{figure*}[!htb]
\begin{center}
\includegraphics[width=8cm]{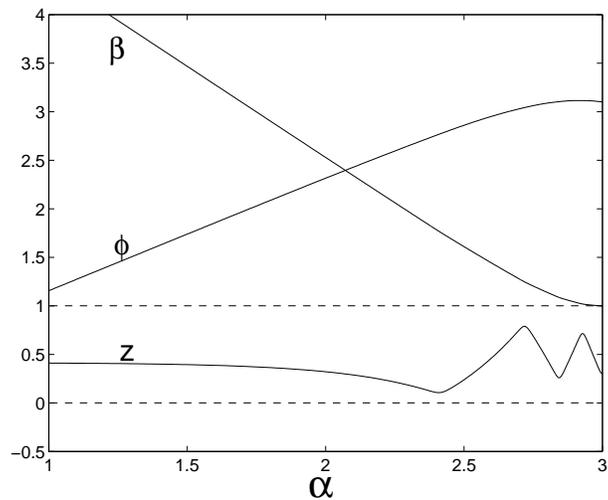}
\caption[fields]{\label{fig:joint} Evolution of the fields $\vphi$, $\beta$ and $z$ for the explicit example given in 
the text. Recall that $z \in [0,1]$ (dashed lines in plot) and that the negative time branch evolution means plots 
read right to left.}
\end{center}
\end{figure*}
\begin{figure*}[!htb]
\begin{center}
\includegraphics[width=8cm]{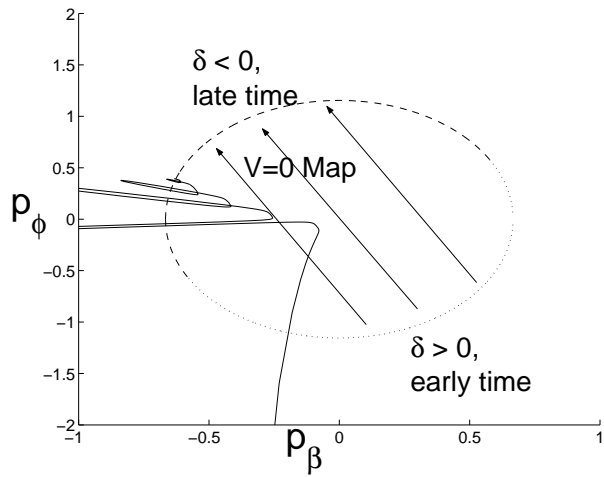}
\caption[RR parameters]{\label{fig:RR} Evolution of the effective RR parameters for the example given in the text 
(solid line). The arrows map the negative time branch evolution between RR solutions in the potential free case, 
see eq.~(\ref{eq:map1}) and the early (late) time part of the RR ellipse, eq.~(\ref{eq:cons1}), are plotted with a 
dotted (dashed) curve.}
\end{center}
\end{figure*}
This is illustrated in figure~\ref{fig:joint}, where we see that  there is no early time RR evolution. The bulk brane 
is moving appreciably and the fields $\vphi$ and $\beta$ start the transition to their late time RR solutions 
straightaway. From this point we see $\vphi$ and $\beta$ continue their transition to the late time solutions while 
$z$ rebounds across the orbifold, prevented from striking the boundary by the membrane potential, until it stabilizes.

We can get a different perspective on this behaviour by considering the effective RR parameters. Figure~\ref{fig:RR} 
shows the ellipse, eq.~(\ref{eq:cons1}), representing early and late time potential free solutions. As expected, in the 
presence of a  potential we start well away from the ellipse, but as $V\rightarrow0$,  $p_\vphi$ and $p_\beta$ return to 
its centre. They then make the transition to the late time half of the ellipse, with their track punctuated by four 
deviations as the potential becomes strong when the bulk brane approaches a boundary.

Note that, although our numerical results generically show this rebounding behaviour, our effective theory does not 
model a collision and is no longer reliable if the bulk brane strikes a boundary, rather than just approaching it and 
rebounding.

\subsection{Interpretation of results}

In all cases we did indeed find that, at late times (i.e. $t\rightarrow
0_-, ~\alpha\rightarrow -\infty$ ), $\vphi$ and $\beta$ followed one of
the $V=0$ late time solutions. 
Perhaps surprisingly, considering the shape of the potential, this was
not necessarily accompanied by late time expanding orbifolds and the 
potential going to zero. 

In cases where the orbifold did contract at late times (parameterised by $p_{\beta}>0$) it was not free to contract as 
rapidly as in the potential free case. 
Using the RR approximation, eq.~(\ref{eq:RR_approx}), in the equations of motion~\footnote{It is simpler to use the 
standard eom in terms of proper time derivatives rather than eqs.~(\ref{eq:jon_eom}).} we find that the potential is 
negligible (even when growing asymptotically) at late times provided $p_{\beta}<(1/3)$. If the orbifold contracts any 
faster the potential acts to slow the contraction until $p_{\beta}<(1/3)$. 
This provides an extra constraint to our late time solutions on top of the $V=0$ late time conditions, i.e. $\phi$ 
and $\beta$ will evolve to the part of the ellipse, eq.~(\ref{eq:cons1}), where $\delta=p_\beta-p_\phi <0$ and 
$p_{\beta}<(1/3)$.

These late time contracting orbifolds are of interest because the potential is no longer exponentially suppressed at 
late time and can grow asymptotically when $p_{\beta}>0$, see eq.~(\ref{eq:RR_approx_pot}). This could provide a way 
of transferring isocurvature perturbations into curvature perturbations in the moving brane scenario, a simplified 
version of this scenario is studied in Ref. \cite{DiMarco:2002eb}. 

Finally, regarding our truncation, one should beware that, in the cases where $p_{\beta}>0$, it is possible that, 
while in the weak coupling regime, we can evolve into the region of parameter space where the axions are no longer 
in a minimum of the potential, forcing us to include all the axionic equations.


\section{NON-PERTURBATIVE EFFECTS AND BACKGROUND FLUID}
\label{sec:yaiza}      
 
In order to let the fields evolve in a cosmologically realistic background, we
 will include a background perfect fluid in the positive time
 branch of our model. 

 We will now look at the solutions of the system governed
 by eq.~(\ref{eq:action}), with the particular scalar potential given in
 eq.~(\ref{eq:scalarpot}). It proves convenient to use the chiral
 superfield notation for the analysis, and then translate our results to the component
 field expressions. 

 We consider the
 simple case of homogeneous, time-dependent, fields in a spatially flat
 FRW space time background, and let them evolve. Given this ansatz, we find the following
equations of motion for the complex superfields
\begin{equation}
\ddot{\Phi}^{i}+3H\dot{\Phi}^i+\Gamma^{i}_{jk}\dot{\Phi}^{j}\\
\dot{\Phi}^{k}+K^{i\bar j}\partial_{\overline{j}}{V}=0 \;\;,
\label{eq:fulleom}
\end{equation}
where $\Phi^i=(S,T,Z)$, $\dot{\Phi}^i=\partial{\Phi}^i/\partial{t}$, $\partial_{\overline{j}}V=\partial{V}/\partial{\overline{\Phi}^{\overline{j}}}$
and the connection on the K\"ahler manifold has the form
\begin{equation}
\Gamma^{n}_{ij}=K^{n\bar l}\frac{\partial K_{j\bar l}}{\partial \Phi^{i}} \; .
\label{eq:gamma}
\end{equation}

\noindent In addition, we obtain the Friedman
equation for the Hubble factor 
$H=\dot{a}/a=\dot{\alpha}$, where
$\alpha(t)={\rm ln} \arrowvert{a(t)}\arrowvert$ was defined in
section \ref{sec:jon_sols} and $a(t)$ is the scale factor of the Universe,
\begin{equation}
3H^2=\kappa_P^2 (\rho_{\phi}+\rho_b) = K_{i\bar j}\dot{\Phi}^{i}\dot{\Phi}^{\bar j}+V+\kappa_P^2\rho_b \;\;,
\label{eq:fullfriedman}
\end{equation}
with $\rho_b$ the energy density of the fluid, which evolves with the scale factor 
in the usual way
\begin{equation}
\rho_b=\rho_{b0}/a^{3(w_b+1)} \;,
\label{eq:rho}
\end{equation}
where $w_b$ defines its equation of state: $p_b=w_b\rho_b$. We
will use radiation, i.e. $w_b=1/3$.

It is worth splitting the equations
of motion for the complex chiral superfields into real and imaginary parts
\begin{equation}
\ddot{\phi}^{i}_{R}+3H\dot{\phi}^{i}_{R}+\frac{1}{2}\Gamma^{i}_{jk}(\dot{\phi}^{j}_R
\dot{\phi}^{k}_R-\dot{\phi}^{j}_I\dot{\phi}^{k}_I)+\frac{1}{2}K^{i\bar j}\partial_{j_R}V=0 \;\;,
\label{eq:reom}
\end{equation}
\begin{equation}
\ddot{\phi}^{i}_I+3H\dot{\phi}^i_I+\frac{1}{2}\Gamma^{i}_{jk}(\dot{\phi}^{j}_I\\
\dot{\phi}^{k}_R+\dot{\phi}^{j}_R\dot{\phi}^{k}_I)+\frac{1}{2}K^{i\bar j}\partial_{j_I}V=0 \;\;,
\label{eq:ieom}
\end{equation}
 where now $\phi^{i}_{R}=(S_R,T_R,Z_R)$ ($\phi^i_I=(S_I,T_I,Z_I)$) refers to the real 
 (imaginary) part of the  superfields;  $\partial_{j_R}$
($\partial_{j_I}$) are
used to denote the derivative of the potential with respect
to the real (imaginary) parts of the fields. The scalar potential $V$ is given, in terms of chiral superfields, in 
eq.~(\ref{eq:scalarpotapp}).
As already mentioned at the beginning of section~ \ref{sec:jon_sols}, the dependence of this 
potential on the axions is contained in the cosine term, and a minimum is found for
$\theta=\pm 2n\pi$, i.e.
\be
Z_I=\frac{1}{2}(T_I\pm \frac{n}{q}) \;\;,
\ee
and no explicit dependence on $S_I$. It seems natural, therefore, to start by solving 
eq.~(\ref{eq:reom}) with the axions fixed at the minimum shown above.

\subsection{Truncated action}

We start off by choosing a particular vacuum for the axions, which will be given by
\begin{eqnarray}
& &S_I=S_{I0} \;\;, \nonumber\\
& &T_I=0 \;\;, \label{eq:axiontrun}\\
& &Z_I=\pm\frac{n}{2q}\nonumber \;.
\end{eqnarray}
Note that the real parts of the fields are bounded by the constraints that define
the field-space region in which the low energy effective theory is
valid, written in eqs.~(\ref{eq:parameterspace1})-(\ref{eq:parameterspace3}).
We restrict our model to the $M<0$ region, where M was defined in
eq.~(\ref{eq:m}), to consistently truncate the action.

\begin{figure*}[!htb]
\begin{center}
\includegraphics[width=8cm]{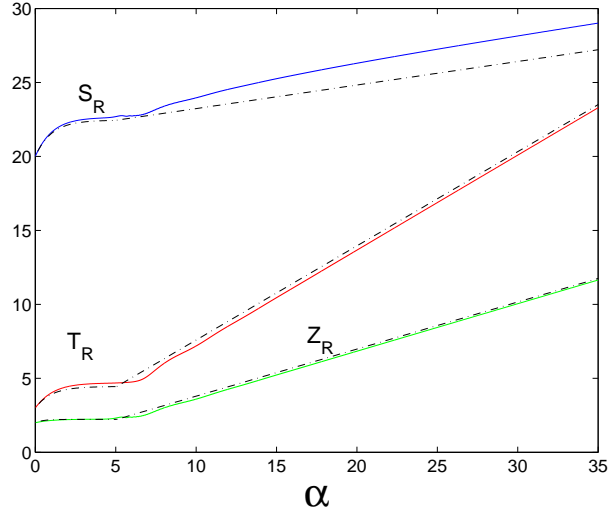}
\caption[moduliev]{\label{moduliev} The real parts of the superfields, \blue{$S_R$}, 
\red{$T_R$}, \green{$Z_R$} as a function of  $\alpha(t)$, which is essentially the number of
 e-folds. The dash-dotted lines show the analytic approximations explained in the text.}
 \end{center}
\end{figure*}

We proceed by evolving the real parts of the superfields in the presence of a background
 fluid that dominates the total energy from the beginning. Under these initial conditions, 
 the fields undergo two different stages, as we can see in figure~\ref{moduliev}.

{\bf 1.} At {\em early times} the fields start evolving because of the initial
velocity they are given,  which is much more important than the
potential energy they have,  until the friction term $3H\dot{\phi}^i$
becomes big enough to freeze them. That is, friction holds the real parts of the superfields on the
potential slope where they remain fixed at a constant value for a while.

\begin{figure*}[!htb]
\begin{center}
\includegraphics[width=8cm]{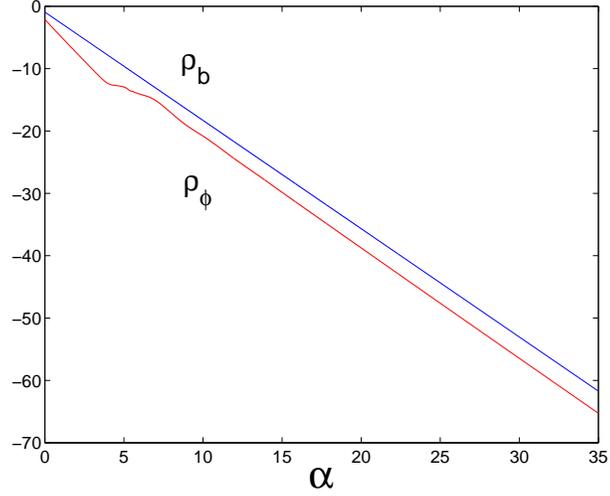}
\caption[fig:endens]{\label{fig:endens} Logarithm of the energy densities for the
 background fluid, \blue{$\rho_b$}, and the real parts of the superfields, \red{$\rho_{\phi}$},
 as a function of $\alpha$.}
 \end{center}
 \end{figure*}

{\bf 2.} As the background fluid energy density, $\rho_b$, decreases with
time, the Hubble factor becomes smaller, and the friction term
is no longer able to hold the fields. The real parts of the superfields then
roll down the potential slope and are driven to an attractor which
makes them scale like the background fluid, as we can see in
figure~\ref{fig:endens}.
Although the fields do
have an attractor solution which makes them mimic the fluid behaviour independently of
their initial conditions, they never dominate the background energy
density. This, as shown in  Ref.~\cite{Ng:2001hs}, rules out any hope for them to become the quintessence field. 

\begin{figure*}[!htb]
\begin{center}
\includegraphics[width=8cm]{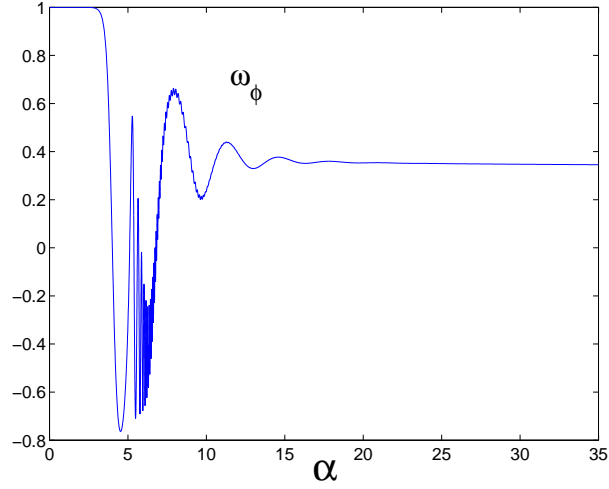}
\caption[fig:wstz]{\label{fig:wstz} Equation of state for the real parts of the superfields as a function 
of $\alpha$.}
\end{center}
\end{figure*}

In figure $\ref{fig:wstz}$ we plot the equation of state
of the real parts of the superfields as a function of  the number of e-folds, showing
that, at late times, they will oscillate around $w=w_b$ ($=1/3$ throughout this paper).
This, again, is
far from the quintessential behaviour, where $-1\leq w_Q\leq-0.7$. 

We can even go further and provide analytic solutions for these two stages.

{\bf 1.}  {\em Early stage}: the evolution of the fields is dominated by the kinetic term, thus 
we can neglect the scalar potential ($V\sim 0$). In this regime we can use the
   complete solution for the fields and their equations of motion
   given in Ref.~\cite{Bastero-Gil:2002hs},  but under the assumption that the
   background fluid dominates in the
   Friedman equation, i.e.
   ${\rho_{kin}}/{\rho_{fluid}} \ll 1$. The early behaviour of
   the fields can then be approximated by
\begin{eqnarray}
& &S_R^e\approx{A_1+A_2e^{-c\alpha}} \;,\nonumber\\
& &T_R^e\approx{B_1+B_2e^{-c\alpha}}\; ,\label{eq:earlyap}\\ 
& &Z_R^e\approx{T_R^e(Z_1+Z_2e^{-c\alpha})}\; .\nonumber
\end{eqnarray}
Therefore, as $\alpha$ grows the second term on the right hand side of these equations
decreases quite fast, and we end up with the constant values that
correspond to the freezing period we mentioned before.
 
{\bf 2.} {\em Late stage}: in the scaling regime the evolution of the fields can be
   approximated by
\begin{eqnarray}
& &S_R^l\approx C_1\alpha+S_R^m\;,\nonumber\\
& &T_R^l\approx C_2\alpha+T_R^m\;,\label{eq:lateap}\\
& &Z_R^l\approx C_3\alpha+Z_R^m\;.\nonumber
\end{eqnarray}
It is also possible to determine the value of these six constants. To do so, we will
perform some approximations in the equations of motion for the real parts of the fields, 
eq.~(\ref{eq:reom}), which, in terms of $\alpha$, is given by
\begin{equation}
\phi^{n''}_R=\phi^{n'}_R(K_{ij}\phi^{i'}_R\phi^{j'}_R-3)-\Gamma^{n}_{ij}\phi^{i'}_R\phi^{j'}_R+\frac{1}{2H^2}[\kappa_{P}^2(1+w_b)\rho_b\phi^{n'}_R-K^{nj}\partial_{j}V] \;\;,
\label{eq:prime}
\end{equation}
where $\phi'=d\phi/d\alpha$. These are the following
\begin{itemize}
\item The background fluid dominates the total energy density, which results in two conditions

\noindent a) Its energy density, $\rho_b$, is much bigger than the kinetic energy of the real parts of the superfields
\begin{equation}
K_{ij}\phi^{i'}_R\phi^{j'}_R \ll 3 \;\;.
\end{equation}
\noindent b) It dominates over the scalar potential
\begin{equation}
V \ll \kappa_{P}^2\rho_{b} \;\;.
\end{equation}

Therefore, the fluid will dominate the Friedman equation
\begin{equation}
3H^2\approx \kappa_{P}^2\rho_{b}\;\;.
\label{eq:friedap}
\end{equation}

\item  The symmetry in eq.~(\ref{eq:csfsym}) for the real parts of the superfields, has a fixed
point at $Z_R=T_R/2$ which defines a minimum of the potential along the $Z_R$ direction. In the
positive time branch case, the presence of the fluid acts as a
friction term that causes the fields to have a kinetic energy of the
same order of magnitude as the potential one. This is why the $Z_R$
modulus always ends up along
the valley that defines the minimum in the $Z_R$ direction.
Physically, this means that the additional bulk 3-brane will always be stabilized
at the middle point in the orbifold direction, i.e. in the middle
between the two boundaries (observable and hidden). In this sense, the
background fluid stops the fields from acquiring
enough kinetic energy to jump out of the valley defined by minimum of
the potential in that direction.

We can therefore
reduce our system to only two real fields by substituting
$Z_R^l=\frac{1}{2}T_R^l$ in our equations.

\item We are in weak coupling regime ($\epsilon \ll 1$), so that
$qT_R \ll S_R$.

\item  The mixing terms are small for the region we are working in, i.e.
$\Gamma^{n}_{ij}\phi^{i'}_R\phi^{j'}_R \ll 3\phi^{n'}_R$.

\item  The slope of the potential in the $T_R$-direction is much steeper than
in the $S_R$-direction as the potential depends exponentially on $T_R$ while
it has a polynomial dependence on $S_R$, therefore
$K^{iT_R}\partial_{T_R}V \gg K^{ij}\partial_{j}V$ for $j\neq T_R$.
\end{itemize}

Equation ($\ref{eq:prime}$) will, after making these approximations, reduce to
\begin{equation}
\phi^{n'}_R\approx \frac{K^{nT_R}\partial_{T_R}V}{3H^2(1-w_b)}{\Bigg\vert}_{Z_R=T_R/2} \;.
\label{eq:reduced}
\end{equation}
From this we can see that the different scaling slopes for the different
real parts of the superfields will just be due to the $K^{pT_R}$ term, which will determine
the following relations
\begin{eqnarray}
K^{S_RT_R}{\Bigg\vert}_{Z_R=T_R/2}=\frac{q}{4}K^{T_RT_R}{\Bigg\vert}_{Z_R=T_R/2}&\Longrightarrow&C_1=\frac{q}{4}C_2\label{eq:c1c2} \;\;, \\
K^{Z_RT_R}{\Bigg\vert}_{Z_R=T_R/2}=\frac{1}{2}K^{T_RT_R}{\Bigg\vert}_{Z_R=T_R/2}&\Longrightarrow&C_3=\frac{1}{2}C_2\label{eq:c3c2} \;\;.
\end{eqnarray}
Now, solving eq.~($\ref{eq:reduced}$) together with eq.~($\ref{eq:friedap}$), we can easily obtain
\begin{eqnarray}
C_1&=&\frac{3}{8\pi}(1+w_b)\label{eq:c} \;\;,\\
C_2&=&\frac{3}{2\pi q}(1+w_b)\nonumber \;\;, \\
C_3&=&\frac{1}{2}C_2\nonumber \;\;,
\end{eqnarray}
and 
\begin{equation}
T_R^{m}\approx \frac{1}{2\pi q}\ln{\frac{64h^2\pi^2q^3}{27(1-w_{b}^2)\rho_{0}\kappa_P^2}} \;.
\label{eq:trm}
\end{equation}
Note that this expression depends on $q$, $\omega_b$ and $\rho_{b0}$, so that $T_R$ will end up having the same 
values independently of the initial conditions for the moduli. This implies that the orbifold size will be, 
at any time, dependent only on the characteristics of the background fluid and on the charge of the moving brane.  

We can now match both stages: using the approximations in both regimes and matching them for $\alpha\gg 1$, we can 
compute the rest of the constants in eqs.~($\ref{eq:earlyap}$) and  ($\ref{eq:lateap}$), and have a complete 
analytic solution given in  Appendix~\ref{sec:anal_approx}, together with an estimate of the time at which the scaling 
regime starts
\begin{equation}
\alpha^{m}\approx \frac{1}{C_2}(B_1-T_R^m)\; .
\label{eq:alpham}
\end{equation}
To complete the late stage approximation, we will write also the matching constant for the $S_R$-field
\begin{equation}
S_R^m \approx \frac{q}{4}(T_R^m-B_1)+A_1\;,
\label{eq:srm}
\end{equation}
where all the remaining constants are given in Appendix~\ref{sec:anal_approx}.
  
As we can see in figure~\ref{moduliev}, the analytic solution reproduces 
accurately the numerical one for the early and late times. There is a
slight discrepancy at late time in our analytic approximation for the 
$S_R^m$, but we can see that the slope of its late time approximation,
$C_1$, is correct.

 We can now express these solutions  in terms of the component
 fields; for the early stage of the evolution we get
\begin{eqnarray}
\vphi_e&\approx&a_1+a_2e^{-c\alpha}\nonumber \;,\\
\beta_e&\approx&b_1+b_2e^{-c\alpha}\label{eq:compearly} \;, \\
z_e&\approx&z_1+z_2e^{-c\alpha}\nonumber \;,
\end{eqnarray}
whereas, at late stages
\begin{eqnarray}
\vphi_l&\approx&c_1\nonumber \;, \\
\beta_l&\approx&\ln{\alpha}+c_2\label{eq:complate} \;, \\
z_l&\approx&\frac{1}{2}\nonumber\;,
\end{eqnarray}
where all the constants are given in Appendix~\ref{sec:anal_approx}.
In figure~\ref{fig:compf} we have plotted the evolution of the component fields
together with the early and late time approximations. Again the
analytical results coincide with the numerical ones.
\begin{figure*}[!htb]
\begin{center}
\includegraphics[width=8cm]{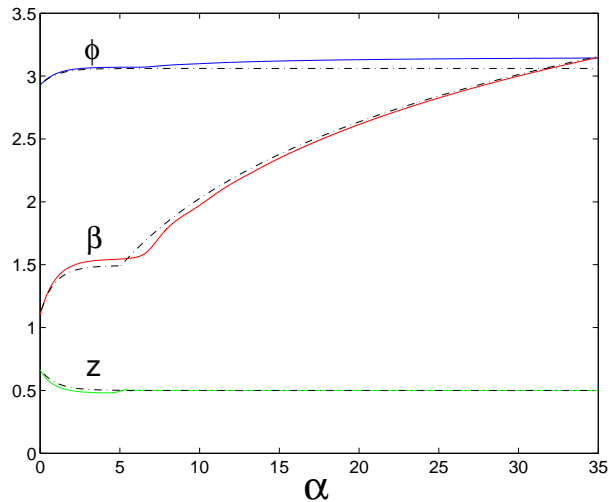}
\caption[fig:compf]{\label{fig:compf} Evolution and analytic approximations
(dotted lines) for the component fields \blue{$\vphi$}, \red{$\beta$}, \green{$z$}  
as a function of $\alpha$.}
\end{center}
\end{figure*}
We can, for completeness, study different initial conditions to the background domination
scenario assumed until now.\\

\noindent {\bf 1. Starting with a moduli dominating energy density}

This will translate into the condition,
\be
\rho_{b0}<\rho_{\phi0} \;\;,
\label{eq:rhofield}
\ee
constrained to $\rho_T=\rho_b+\rho_{\phi}<M_G^4$, where $M_G$ defines the GUT scale. The real parts of the superfields 
will, in this case,  evolve quite fast, but the friction term due to the
fluid will slow them down so that, at some point, the field energy
density will get smaller than the background one. Thus the fields will
either grow too fast to stay in the weak coupling regime, or follow
the behaviour explained before, i.e. once their density becomes
smaller than $\rho_b$ they will end up scaling with the background energy density. 
The only difference
will now be that we cannot use the early approximations anymore.\\

\noindent {\bf 2. Starting with a large kinetic energy for $Z_R$}

When we give  the $Z_R$-field a large  enough initial velocity, the 3-brane
will move towards the boundary and hit it, bouncing back again,
causing our theory to
break down. There are certain cases in which the
3-brane will just get close to the boundary, go up the
potential slope and bounce back. This
will affect the evolution of the other real parts of the superfields because of their
mixing terms in the equations of motion. In
the case of  $S_R$, which is coupled to $Z_R$ in quite a strong
way 
through $\Gamma^{S_R}_{Z_RZ_R}$ as well as
$\Gamma^{S_R}_{S_RZ_R}$, the fact that $Z_R$ bounces back will drag it along,
so that the volume of the Calabi-Yau  will become smaller. Meanwhile the $T_R$-modulus will
not be affected by the other fields'  behaviour,
as in its equation of motion the dominant mixing
terms is by far  $\Gamma^{T_R}_{T_RT_R}$.
The decrease in the value of $S_R$, while  $T_R$ is still increasing, will lead the system to
strong coupling (that is $S_R \gg qT_R$ implying $\epsilon \gg 1$), so that our model breaks down and results are not
valid anymore.

\subsection{Including axions}
\begin{figure*}[!htb]
\begin{center}
\includegraphics[width=8cm]{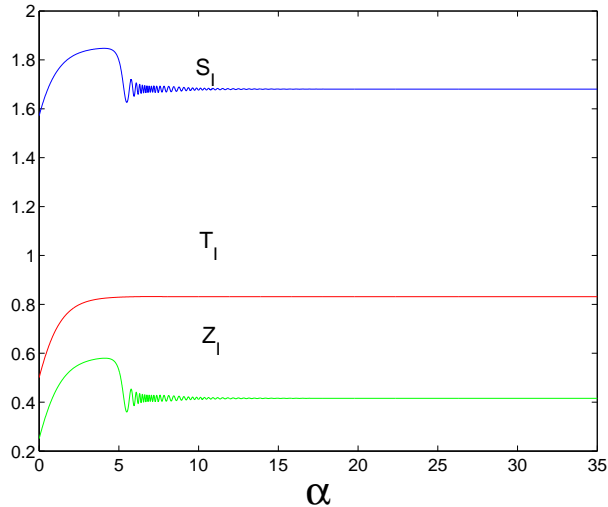}
\caption[fig:axions]{\label{fig:axions} Evolution of the axions, \blue{$S_I$}, \red{$T_I$}, \green{$Z_I$} as a 
function of the number of e-folds, $\alpha$.}
\end{center}
\end{figure*}

If we let the axions evolve as well, we will see that their behaviour is similar to that of the real parts:
they will just evolve while their kinetic
energy is bigger than the friction term due to the background fluid
and then, they will slow down and be driven towards their minimum where they will stabilize at their minima after 
some oscillations, as we can see in figure~\ref{fig:axions}. The evolution of the real parts of the superfields will 
not be affected by the axionic one as the axions only appear in the real parts of the
superfields equations of motion, eq.~(\ref{eq:reom}), through the mixing term, which is typically very small. Only 
in certain cases, in which the axions have a large amount of initial kinetic energy, will the real parts of the 
superfields modify their behaviour.
Because this mixing term has a negative sign, the real parts will decrease while the axions
still have an effect on the evolution, until the friction term dominates again pushing the
axions to their minimum, and letting the fields go to the same scaling
regime again. As before, the slopes of this late approximation are really accurate.

\subsection{Interpretation of results}

There are four main results we would like to highlight:

{\bf 1.} The axions, in general, will not affect the real parts of the superfields evolution,
as they will always stabilize at the minimum of the potential in the
axionic direction, thanks to the damping of the background fluid.
Thus, when we consistently truncate the action to its minimal version, where the axions 
rest at their minimum, with certain constant values given in
eq.~(\ref{eq:axiontrun}), we will not lose any important features. Only if their initial kinetic
energy is large will they
have an effect on the real parts of the superfields evolution, but just through the early
stages, leading to the same scaling solution at late time. 

{\bf 2.} The additional bulk 3-brane gets stabilized at the middle
point between the two boundaries, which is shown in figure~\ref{fig:compf} by the $z$ field
going always to the value $z=0.5$. This behaviour is due to the
symmetries in our potential, eq.~(\ref{eq:csfsym}), together with the fact that there is
a background fluid that dominates the total energy, preventing the
kinetic energy of the fields from growing over the potential one.

{\bf 3.} The orbifold size, $\pi\rho T_R$, grows linearly with the
number of e-folds, eq.~(\ref{eq:lateap}),
with a speed, $C_2$, that depends only on the charge of the moving brane and
the background fluid properties, as shown in eq.~(\ref{eq:c}). Thus, its
size will be independent of the initial conditions of the chiral superfields.

{\bf 4.} Our model always tends to strong coupling regime at late
times. This was suggested in earlier
papers in which the moving 3-brane was free, as there was neither a
background fluid nor an instanton superpotential, see Ref.~\cite{Copeland:2001zp},  so we can conclude that the 
addition of the
membrane instantons does not affect that result. 
We can, in this case, understand why this happens: it is due to the slope of the $T_R$-modulus ($C_2$) being bigger than
the one of the $S_R$-field ($C_1$), which comes from the fact that $K^{T_RT_R}$
is bigger than $K^{S_RT_R}$ in this regime where $Z_R=T_R/2$. 

{\bf 5.} The volume of the Calabi-Yau, $\phi$, tends to a constant at
late time, as we can see in eq.~($\ref{eq:complate}$). This is
due to the relation between the scaling regime slopes $C_1=(\pi/4)C_2$,
that will cancel the time ($\alpha$) dependence in $\phi_{l}$
\begin{equation}
\phi_{l}=\ln{(S_R^{l}-q{Z_R^l}^2/T_R^l)}=\ln{(S_R^{m}-\frac{q}{4}T_R^{m})}\;\;,
\end{equation}
but this will happen when going to strong coupling, where our theory
breaks down.


\section{CONCLUSIONS}
\label{sec:conclusions}

In this paper we have explored further aspects of low energy heterotic
M-theory. We have presented new symmetries of the full action
and used these to generate a new class of solutions with non-trivial axion dynamics. We saw that the non-axionic 
fields have the same evolution while two of the axions now make a similar transition to the bulk brane, from one 
constant value to another, leaving the system to return to a two field rolling radii model asymptotically.  The 
similarity of these solutions to certain PBB ones suggests the axions will produce a scale invariant isocurvature 
perturbation.

We then investigated the effect of membrane instantons on the dynamics of the system, both with and without a 
background fluid. In the absence of a background fluid we find that, despite a more complicated evolution, the 
Calabi-Yau and orbifold moduli behave as rolling radii in the late time limit with the potential dropping out of 
the dynamics.  During the evolution to these late time solutions we find the membrane potential attempts to prevent 
the bulk brane striking a boundary, leaving it to stabilize in the bulk. Additionally, although negligible in the 
background equations of motion, the potential can grow asymptotically at late times, providing a possible mechanism 
for transferring isocurvature perturbations into curvature perturbations.

We then included a perfect fluid in the background for the positive time
branch of the evolution of the Universe. This fluid acts as a friction term for the fields evolution,
stopping their kinetic energy from growing bigger than the potential
one. This has an important effect in

\noindent i) The axions which, after evolving, stabilize at the minimum of the
potential in their direction. They drop out of the real part
of the
fields equation of motion (\ref{eq:reom}) and, while evolving, do  not change the results
found for the truncated version of the action.

\noindent ii) The moving brane, that is driven to the minimum of the
potential in its direction, that is, it is stabilized at the
middle point between the orbifold boundaries.

In this case, the real parts of the superfields scale with the background fluid at late times. 
This is a nice feature, typical of any quintessential model but, in this case, it is insufficient to
explain the amount of dark energy in the Universe, as the moduli energy density never 
dominates over the background fluid one. We were able to find  a set of analytic solutions,
given by eq.~(\ref{eq:lateap}) for all the fields. In general $S_R$, $T_R$ and $Z_R$ increase linearly with 
$\alpha$, with a speed, eq.~(\ref{eq:c}), that depends only on the charge of the 3-brane, 
$q$, and the background fluid characteristics, $\rho_B$, and which is independent of their initial conditions. 

Because of this scaling behaviour, the orbifold size, $\pi\rho T_R$, ends up being always the same
at a certain time, given a fixed background fluid and charge, as we can see in eq.~(\ref{eq:trm}). But the different 
slopes of the scaling solutions for the different fields 
drive the system to strong coupling at late times.

In the future, we will investigate two possible directions. Firstly, we
will try to find the higher dimensional fields that will act as
a background fluid in our 4-dimensional effective theory and see the
changes this might have in the superfields evolution. Secondly, we
will  be calculating the perturbation spectra of two of the new
classes of solutions, i.e. the new axionic solutions and the solutions
with a potential but without a fluid, to see whether either can
generate scale invariant perturbations. We know that the minimal, potential free, solutions~\cite{Copeland:2001zp} 
cannot generate either adiabatic or
isocurvature perturbations with a scale invariant spectrum Ref.~\cite{DiMarco:2002eb}.  


\begin{acknowledgments}
The authors wish to thank Andr\'e Lukas and Ed Copeland for very useful
discussions. YS thanks Nelson Nunes for his help. BdC and JR are supported by PPARC and YS by the University of Sussex.
\end{acknowledgments}


\appendix

\section{K\"ahler matrix}
\label{sec:Kahler}
The K\"ahler
matrix,
$K_{i\bar
  j}=\frac{\partial^{2}K}{\partial\Phi^{i}\partial\bar\Phi^{\bar j}}$,
takes the specific form shown below when using the tree level K\"ahler potential, eq.~(\ref{eq:kahler}), 
%
\[{K_{i\bar j}} =  \frac{1}{y^2}\left( \begin{array}{cccc} {1} &
  {\frac{qZ_R^2}{T_R^2}} & {-\frac{2qZ_R}{T_R}} \\
  {\frac{qZ_R^2}{T_R^2}} & {\frac{1}{T_R^2}(\frac{q^2Z_R^4}{T_R^2}+y\frac{qZ_R^2}{T_R}+\frac{3}{4}y^2)} &
  {-\frac{qZ_R}{T_R^2}(y+\frac{2qZ_R^2}{T_R})} \\ 
{-\frac{2qZ_R}{T_R}} & {-\frac{qZ_R}{T_R^2}(y+\frac{2qZ_R^2}{T_R})}
 & {\frac{q}{T_R}(y+\frac{4qZ_R^2}{T_R})}\end{array} \right) \]\label{eq:kahlermatrix}
%
\noindent where $y=2S_R-2q\frac{Z_R^2}{T_R}$. Note that this matrix is a function of
the real parts of the fields only and that it is symmetric.

\section{Scalar potential}
\label{sec:app_pot}

The D=4, N=1, supergravity scalar potential, eq.~(\ref{eq:potential}),
for this particular K\"ahler matrix but for a general superpotential,
is of the form
\be
V=\frac{1}{8y^3T_R^3}\left[|yW_S-W|^2+\frac{4qy}{T_R}\Bigg\vert
  Z_RW_S+\frac{T_R}{2q}W_Z\Bigg\vert ^2+\frac{4}{3}\Bigg\vert
  T_RW_T+Z_RW_Z+\frac{qZ_R^2}{T_R}W_S-\frac{3}{2}W\Bigg\vert^2-3|W|^2\right]\label{eq:potgralsup}\;\;,
\ee
where $y$ has been defined above and $W_i=\partial W/\partial\Phi^i$. 

When we include the specific superpotential of eq.~(\ref{eq:superp}), 
as it was already presented in the text, the scalar potential in terms of the chiral superfields 
can be written as 
\begin{equation}
V=V_{0}(S_R,T_R,Z_R)[A(S_R,T_R,Z_R)+B(S_R,T_R,Z_R)\cos\theta(T_I,Z_I)] \;\;,
\label{eq:scalarpotapp}
\end{equation}
where
\begin{eqnarray}
V_{0}(S_R,T_R,Z_R)&=&-\frac{h^2}{48(S_RT_R^3-qT_R^2Z_R^2)} \;\;, \nonumber\\
A(S_R,T_R,Z_R)&=&(-3-24\pi^2qS_RT_R+8\pi^2q^2Z_R^2-24\pi qZ_R) {\rm e}^{-4\pi
    qZ_R} \nonumber\\ 
& + &(-3-24\pi^2qS_RT_R+8\pi^2q^2Z_R^2-16\pi^2q^2T_R^2\nonumber\\
& + &32\pi^2q^2T_RZ_R-24\pi qT_R+24\pi qZ_R) {\rm e}^{4\pi q(Z_R-T_R)} \;\;, \\
B(S_R,T_R,Z_R)&=&2(-3+24\pi^2qS_RT_R-8\pi^2q^2Z_R^2-16\pi^2q^2T_RZ_R-12\pi
qT_R) {\rm e}^{-2\pi qT_R} \;\;, \nonumber\\
\theta(T_I,Z_I)&=&2\pi q(2Z_I-T_I)\nonumber \;\;.
\end{eqnarray}
Inserting the definition for the component fields given by eq.~(\ref{eq:compfields}) we have
\ba
V_{0}(\phi ,\beta ,z)&=&\frac{-h^{2}}{48e^{3\beta +\phi}}  \;\;, \nonumber \\
A(\phi ,\beta ,z)&=& (-3-24\pi^{2}qe^{\beta +\phi} -
     16\pi^{2}q^{2}e^{2\beta}(z-1)^{2} + 24\pi qe^{\beta}(z-1)) {\rm e}^{4\pi qe^{\beta}(z-1)} \nonumber \\
 &+  &    (-3-24\pi^{2}qe^{\beta+\phi }- 16\pi^{2}q^{2}e^{2\beta}(z)^{2}- 
     24\pi qe^{\beta}z)e^{-4\pi qe^{\beta}z} \; \;, \label{eq:real_pot} \\ 
B(\phi ,\beta ,z)&=& (-6+48\pi^{2}qe^{\beta+\phi}+ 32\pi^{2}q^{2}e^{2\beta}z(z-1)-24\pi qe^{\beta})e^{-2\pi qe^{\beta}} \;\;, \nonumber \\
\theta(\chi,\zeta,z)&=&2\pi q(-4\zeta+4z\chi-2\chi) \nonumber \;\;.
\ea

\section{Analytic solutions}
\label{sec:anal_approx}

The different constants appearing in the early time approximations for $S_R$, $T_R$ and $Z_R$, eqs.~(\ref{eq:earlyap}), 
are given by
\begin{eqnarray} 
A_1&=& S_R(0)-A_2 \;\;, \nonumber \\
A_2&=&\frac{S_R(0)'}{c} \;\;, \nonumber\\
B_1&=& T_R(0)-B_2 \;\;, \nonumber\\
B_2&=&\frac{-T_R(0)'}{c} \;\;, \nonumber\\
Z_1&=&\frac{1}{2} \label{eq:modapp} \;\;, \\
Z_2&=&\frac{Z_R(0)}{B_1+B_2}-Z_1 \;\;, \nonumber\\
c&=& \frac{3}{2}(1-w_b) \;\;. \nonumber
\end{eqnarray}
The analogous expressions in terms of the component fields 
$\phi$, $\beta$, $\gamma$, see eqs.~(\ref{eq:compearly}) are
\begin{eqnarray} 
a_1&=& \ln{(S_R^m-q\frac{{Z_R^m}^2}{T_R^m})}-a_2 \;\;, \nonumber \\
a_2&=&\frac{S-R^(0)'{T_R^m}^2+qZ_R^m(Z_R^mT_R(0)'-2Z_R(0)'T_R^m)}{-cT_R^m(S_R^mT_R^m-q{Z_R^m}^2)} \;\;, \nonumber\\
b_1&=&\ln{T_R^m}-b_2 \;\;, \nonumber\\
b_2&=&\frac{-T_R(0)'}{cT_R^m} \;\;, \label{eq:compfapp}\\
z_1&=&\frac{Z_R^m}{T_R^m}-z_2 \;\;, \nonumber\\
z_2&=&\frac{Z_R^mT_R(0)'-Z_R(0)'T_R^m}{c{T_R^m}^2} \;\;.\nonumber
\end{eqnarray}
Finally, the constants for the late time approximations in terms of the component fields, see
eqs.~(\ref{eq:complate}) are
\begin{eqnarray}
c_1&=&\ln{(S_R^m-\frac{q}{4}T_R^m)} \;\;, \nonumber\\
c_2&=&\ln{C_2} \label{eq:complateapp} \;\;. 
\end{eqnarray}



\begin{thebibliography}{99}
%



\bibitem{Lukas:1998yy}
A.~Lukas, B.~A.~Ovrut, K.~S.~Stelle and D.~Waldram, Phys. Rev. {\bf D59}, 086001 (1999),
hep-th/9803235.
%
\bibitem{Lukas:1998qs}
A.~Lukas, B.~A.~Ovrut and D.~Waldram, Phys. Rev. {\bf D60}, 086001 (1999),
hep-th/9806022.
%
\bibitem{Randall:1999vf}
L.~Randall and R.~Sundrum, Phys. Rev. Lett. {\bf 83}, 4690 (1999), hep-th/9906064.
%
\bibitem{Brandle:2000qp}
M.~Brandle, A.~Lukas and B.~A.~Ovrut, Phys. Rev. {\bf D63}, 026003 (2001),
hep-th/0003256.
%
\bibitem{Dvali:1998pa}
G.~R.~Dvali and S.~H.~H.~Tye, Phys. Lett. {\bf B450}, 72 (1999), hep-ph/9812483.
%
\bibitem{Kehagias:1999vr}
A.~Kehagias and E.~Kiritsis, JHEP {\bf 11}, 022 (1999), hep-th/9910174.
%
\bibitem{Khoury:2001wf}
J.~Khoury, B.~A.~Ovrut, P.~J.~Steinhardt and N.~Turok, Phys. Rev. {\bf D64}, 123522 (2001),
hep-th/0103239.
%
\bibitem{Alexander:2001ks}
S.~H.~S.~Alexander, Phys. Rev. {\bf D65}, 023507 (2002), hep-th/0105032.
%
\bibitem{Burgess:2001fx}
C.~P.~Burgess et al., JHEP {\bf 07}, 047 (2001), hep-th/0105204.
%
\bibitem{Kallosh:2001du}
R.~Kallosh, L.~Kofman, A.~D.~Linde and A.~A.~Tseytlin, Phys. Rev. {\bf D64}, 123524 (2001),
hep-th/0106241.
%
\bibitem{Copeland:2001zp}
E.~J.~Copeland, J.~Gray and A.~Lukas, Phys. Rev. {\bf D64}, 126003 (2001),
hep-th/0106285.
%
\bibitem{Steinhardt:2001vw}
P.~J.~Steinhardt and N.~Turok, hep-th/0111030.
%
\bibitem{Copeland:2002fv}
E.~J.~Copeland, J.~Gray, A.~Lukas and D.~Skinner, Phys. Rev. {\bf D66}, 124007 (2002),
hep-th/0207281.
%
\bibitem{Horava:1996qa}
P.~Horava and E.~Witten, Nucl. Phys. {\bf B460}, 506 (1996), hep-th/9510209.
%
\bibitem{Horava:1996ma}
P.~Horava and E.~Witten, Nucl. Phys. {\bf B475}, 94 (1996), hep-th/9603142.
%
\bibitem{Witten:1996mz}
E.~Witten, Nucl. Phys. {\bf B471}, 135 (1996), hep-th/9602070.
%
\bibitem{Horava:1996vs}
P.~Horava, Phys.  Rev. {\bf D54}, 7561 (1996), hep-th/9608019.
%
\bibitem{Lukas:1998fg}
A.~Lukas, B.~A.~Ovrut and D.~Waldram, Nucl. Phys. {\bf B532}, 43 (1998), hep-th/9710208.
%
\bibitem{Lukas:1998hk}
A.~Lukas, B.~A.~Ovrut and D.~Waldram, Phys. Rev. {\bf D59}, 106005 (1999),
hep-th/9808101.
%
\bibitem{Ellis:1998dh}
J.~R.~Ellis, Z.~Lalak, S.~Pokorski and W.~Pokorski, Nucl. Phys. {\bf B540}, 149 (1999),
hep-ph/9805377.
%
\bibitem{Lukas:1998tt}
A.~Lukas, B.~A.~Ovrut, K.~S.~Stelle and D.~Waldram, Nucl. Phys. {\bf B552}, 246 (1999),
hep-th/9806051.
%
\bibitem{Brandle:2001ts}
M.~Brandle and A.~Lukas, Phys. Rev. {\bf D65}, 064024 (2002),
hep-th/0109173.
%
\bibitem{Derendinger:2000gy}
J.~P.~Derendinger and R.~Sauser, Nucl. Phys. {\bf B598}, 87 (2001), hep-th/0009054.
%
\bibitem{Mueller:1990rr}
M.~Mueller,  Nucl. Phys. {\bf B337}, 37 (1990).
%
\bibitem{Becker:1995pot}
K.~Becker, M.~Becker and A.~Strominger,  Nucl. Phys. {\bf B456}, 130 (1995), hep-th/9507158.
%
\bibitem{Lukas:1999kt}
A.~Lukas, B.~A.~Ovrut and D.~Waldram, JHEP {\bf 04}, 009 (1999), hep-th/9901017.
%
\bibitem{Harvey:1999as}
J.~A.~Harvey and G.~W.~Moore, hep-th/9907026.
%
\bibitem{Moore:2000fs}
G.~W.~Moore, G.~Peradze and N.~Saulina, Nucl. Phys. {\bf B607}, 117 (2001), hep-th/0012104.
%
\bibitem{Lima:2001jc}
E.~Lima, B.~A.~Ovrut, J.~Park and R.~Reinbacher, Nucl. Phys. {\bf B614}, 117 (2001), hep-th/0101049.
%
\bibitem{Lima:2001nh}
E.~Lima, B.~A.~Ovrut and J.~Park, Nucl. Phys. {\bf B626}, 113 (2002), hep-th/0102046.
%
\bibitem{Curio:2001qi}
G.~Curio and A.~Krause, Nucl.\ Phys.\ B {\bf 643}, 131 (2002), hep-th/0108220.
%
\bibitem{DiMarco:2002eb}
 F.~Di~Marco, F.~Finelli and R.~Brandenberger, Phys. Rev. {\bf D67}, 063512 (2003),astro-ph/0211276.
%
\bibitem{Lukas:1999nh}
A.~Lukas and K.~S.~Stelle, JHEP {\bf 01}, 010 (2000), hep-th/9911156.
%
\bibitem{Ng:2001hs}
S.~C.~C.~Ng, N.~J.~Nunes and F.~Rosati, Phys. Rev {\bf D64}, 083510 (2001), astro-ph/0107321.
%
\bibitem{Copeland:1994vi}
E.~J.~Copeland, A.~Lahiri, D.~ Wands, Phys. Rev. {\bf D50}, 4868 (1994), hep-th/9406216.
%
\bibitem{Gasperini:1993em}
M.~Gasperini and G.~Veneziano, Astropart. Phys. {\bf 1}, 317 (1993), hep-th/9211021.
%
\bibitem{Lidsey:1999mc}
J.~E.~Lidsey, D.~ Wands, E.~J.~Copeland, Phys. Rept.{\bf 337}, 343 (2000), hep-th/9909061.
%
\bibitem{Bastero-Gil:2002hs}
M.~Bastero-Gil, E.~J.~Copeland, J.~Gray, A.~Lukas and M.~Plumacher, Phys. Rev. {\bf D66}, 066005 (2002), hep-th/0201040.

\end{thebibliography}
\end{document}